# Ideal Solar Cell Efficiencies

Tom Markvart

Solar Energy Laboratory, Faculty of Engineering and Physical Sciences, University of Southampton, UK, and

Centre for Advanced Photovoltaics, Faculty of Electrical Engineering, Czech Technical University in Prague, Czech Republic

In a recent paper, Guillemoles *et al*[1] attempt to clarify and explain the often cited paper by Shockley and Queisser[2] (SQ) which defines the limits to photovoltaic conversion by a single-junction solar cell. The SQ paper is not easy to read and is therefore easily misunderstood. As modern solar cells approach theoretical efficiency limits, the fundamentals become particularly important and the effort by Guillemoles *et al* is therefore to be welcome. However, in doing so, the authors have fallen into several pitfalls and the aim of the present note is to clarify a number of misconceptions and correct some errors in that paper.

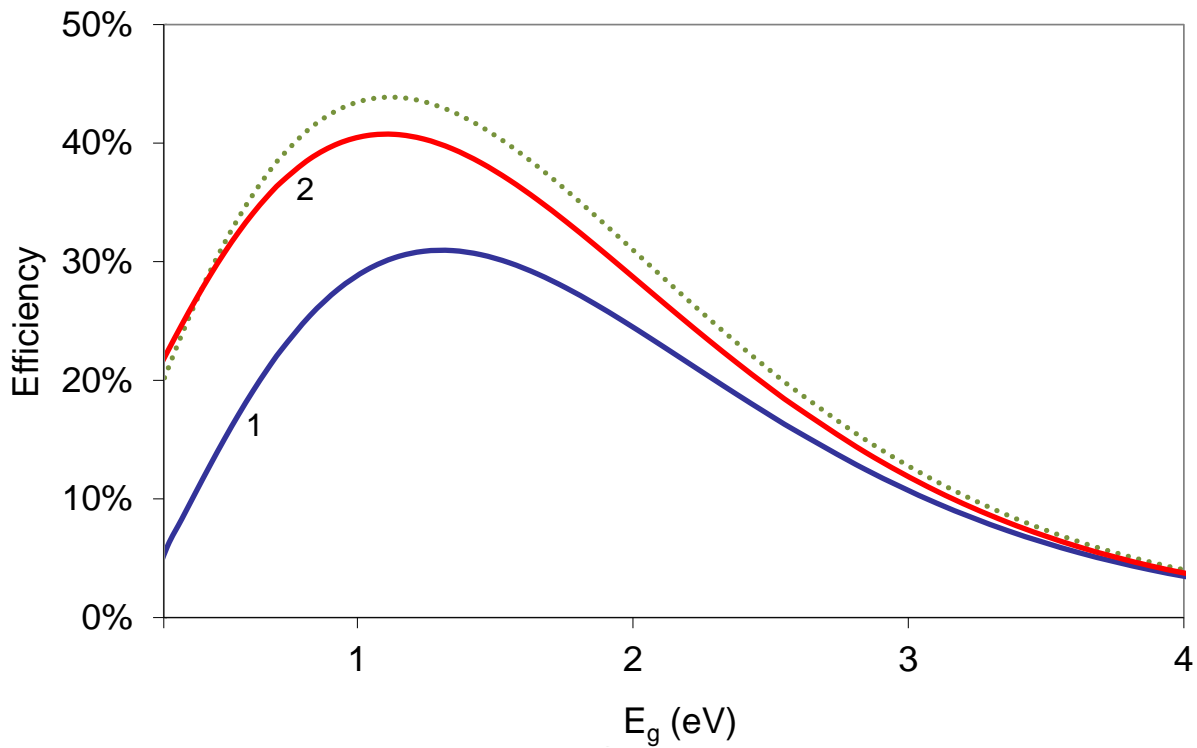

Fig. 1. The frequently quoted Shockley-Queisser efficiency curves: 1 – under "one-sun" illumination, and 2 – under sunlight at maximum concentration ratio of approximately 46,000. Also shown, by the dotted line, is the Trivich-Flinn efficiency (1). Sunlight modelled as black-body radiation at 6000K, cell temperature 300K.

Shockley and Queisser described their result as detailed balance limit which - in an intuitive manner - describes a balance between the incident and emitted photon fluxes rather than a similar thermodynamic term of detailed balancing (also referred to as microscopic reversibility[3]), more akin to the earlier paper by van Roosbroeck and Shockley[4] and more recent reciprocity theorems (see e.g. Rau[5]). The key principle of the paper that the maximum efficiency of a solar cell depends solely on the photon fluxes of the incident and emitted radiation provided the impetus for a later full thermodynamic interpretation,[6] underpinning the earlier SQ suggestion (but see also ref. 7 for an earlier thermodynamic theory of PV conversion). SQ did not find it easy to publish their ground-breaking result (see for example, ref. 8 ). The paper includes more material that is now usually cited but the curves that have survived the test of time are shown in Fig. 1.

Guillemoles *et al* discuss only the SQ curve that corresponds to one sun illumination. This leads them to ascribe – incorrectly – a major part of voltage from the "ideal" value of $E_g/q$ to electrical work of transferring a charge carrier between the contacts (labelled as "isothermal losses" in ref. 1). In fact, the largest part of this loss is of optical nature, a fundamental and unavoidable loss contained in the SQ theory and visible in Fig. 1 as the difference between the maximum-concentration and one-sun efficiencies. Equal to 0.28 V this voltage loss, sometimes called optical entropy generation,[9] and due to the étendue expansion between the incident to emitted beam. In the SQ paper, a similar effect gives rise also to a voltage reduction by $kT_c \ln 2$ due to photon emission from two faces of the solar cell, an effect discussed in more detail in ref. 14.

In identifying the "real" voltage losses relative to the SQ value in Eq. (3), Guillemoles *et al* define somewhat novel figures of merit rather than use more conventional parameters (see e.g. refs. 10, 11, 12). Doing so leads to some confusion and an incorrect estimate of one of these parameters ($F_{em}$), usually defined in terms of emitted photon fluxes rather than via the dark saturation current of the solar cell. Indeed, if $J_o^{real} = J_o^{QE} / Q_e^{lum}$, as defined in Table 1, where $Q_e^{lum}$ is the efficiency of luminescent emission by the solar cell when operating as a light-emitting diode then, by optoelectronic reciprocity[5], $J_o^{QE} = EQE_o J_o^{SQ}$, where $EQE_o$ is the external quantum efficiency for the emitted light. It therefore follows that the parameter $F_{em}$ of ref. 1 is just the reciprocal of $EQE_o$ and never smaller than unity, contrary to the claim on p.504. This conclusion can be easily understood in physical terms: because of lower efficiency, a real solar cell will always emit a photon flux no higher than an ideal (SQ) cell, and therefore $F_{em} \geq 1$ - in contrast to the higher dark saturation current obtained by dividing with $Q_e^{lum}$. It is also worth noting that this result is a consequence of both potentially lower-than-unit emissivity at the emission wavelengths (losses in Stage A, as attributed in ref. 1), but also due to losses in carrier transport to junction, and therefore originating from losses in Stage C.

Possibly more subtle but nevertheless important from the fundamental viewpoint is the fact that the construction in Fig. 2b is not due to Shockley and Queisser, as least not what is generally understood to be their acclaimed work leading to the efficiencies shown by the full lines in Fig. 1. In fact, the maximum efficiency implied in Fig. 2b, as given by

$$\eta_{TF} = E_g \int_{E_g}^{\infty} \Phi(E)dE \bigg/ \int_{0}^{\infty} E\Phi(E)dE \qquad (1)$$

is due to Trivich and Flinn,[13] published some six years before the paper by Shockley and Queisser. Shockley and Queisser reproduce this efficiency under the name of "ultimate efficiency", shown by the dotted line in Fig. 1. The construction in Fig. 2b is due to Henry[14] who discussed the full complexity of obtaining measured solar cell parameters by this technique and whose construction clearly highlights the difference between the "maximum concentration" and " one sun" efficiencies discussed earlier in this note. The use of TF efficiency (1) in place of the true SQ efficiency results in an error (which seems to have disappeared from Fig. 2a), already contained in ref. 15, where the Trivich-Flinn efficiency is also incorrectly used as the starting point.